# A theoretic model for sonogenetic antiarrhythmia

**Jing-Jing Wang[1], Xiang Gao[1,2*]**

[1]School of Physics and Information Technology, Shaanxi Normal University, Xi'an, China

[2]Max-Planck-Institute for Dynamics and Self-Organization, Göttingen, Germany

**\* Correspondence:**
Corresponding Author: Xiang Gao
E-mail:gaoxiang.gnaixoag@gmail.com

**Abstract**

Sonogenetics can be used as a new alternative for treating arrhythmia due to its advantages of noninvasive, high safety and strong penetration. In the treatment of arrhythmias by sonogenetics, cardiac myocytes are deformed by ultrasonic radiation force. We quantitatively calculated the shape variation of cardiomyocytes under ultrasonic radiation force, and the deformation of cardiomyocytes caused the change of membrane tension. Membrane tension consists of two parts, plasma membrane tension and cortical tension between the cell membrane and cytoskeleton. Since plasma membrane tension was mainly considered in existing experiments, we proposed a quantitative model of the relationship between ultrasonic radiation force and plasma membrane tension. The Boltzmann relationship between plasma membrane tension and ion channel opening probability is presented based on the experimental results of ion channel activation by stretching. Finally, a quantitative model was obtained for ultrasonic radiation force to regulate the opening probability of ion channel activated by stretching. Based on this quantitative model, we proposed the regulation mechanism of ultrasonic radiation force under hypercompression and hyperstretching, and verified that this mechanism can eliminate arrhythmias by sonogenetics.

## 1 Introduction

In 2015, Ibsen et al[1] proposed the concept of sonogenetics. Sonogenetics works by using genetic engineering techniques to infect a target cell with stretch activated ion channels (SAC). According to the experimental results of electron microscopy[2][3], SAC channels were opened under ultrasonic radiation force[4]. Ions flow along their electrochemical gradient through SAC channels, resulting in changes in cell transmembrane potential [5].Cells are activated or inhibited with the change of membrane potential[6][7], thus regulating the behavior of cells. As a new cell regulation method, sonogenetics has attracted extensive attention in many fields, especially in the field of neuroscience, such as neural cell regulation[8], regulation of target gene expression [9] and regulation of tumor cell death[10].If sonogenetics is applied to cardiac myocytes, the excitability of cardiac myocytes can be regulated by ultrasonic radiation force, so as to eliminate abnormal electrical signals (spiral waves and turbulence) on the heart noninvasion, high safety and strong penetration[11][12][13], so as to achieve the effect of treating arrhythmias. Our previous work demonstrated that sonogenetics can be used to treat arrhythmias[14], but in this process, it is necessary to consider how ultrasonic radiation forces control the opening probability of stretching-activated ion channels. In this paper, the ultrasonic radiation force causes the deformation of myocardial cells. We quantitatively calculate the deformation of myocardial cells under the ultrasonic radiation force. The deformation of myocardial cells causes the change of membrane tension. Membrane tension consists of two parts, plasma membrane tension and cortical tension between the cell membrane and cytoskeleton. Since plasma membrane tension was mainly considered in existing experiments, we proposed a quantitative model of the relationship between ultrasonic radiation force and plasma membrane tension. According to the experimental results of ion channel activation by stretching, the Boltzmann relationship between plasma membrane tension and ion channel opening probability was presented[15][16][17]. Solving the problem of opening probability of stretch activated ion channel controlled by ultrasonic radiation force is a key step to promote the application of sonogenetics in clinical treatment of arrhythmia. This method can also be helpful for studying the non-invasive regulation of neural cell action potential by sonogenetics .

## 2    Method

In the experiment of treating arrhythmia with sonogenetics, we applied the ultrasonic radiation force vertically to the myocardial tissue along the direction of contraction and relaxation of myocardial cells, so that the myocardial tissue would be deformed under the action of ultrasonic radiation force. Myocardial tissue is composed of a series of myocardial cells, and when the tissue is deformed, each cell is also deformed with the same shape variable[18]. If ultrasonic radiation force causes myocardial tissue to stretch or shorten by 10%, we assume that each of its cardiac cells also extends or shortens by 10%. Although the cell is a viscoelastic material, we mainly consider the influence of external force, namely ultrasonic radiation force. The cell is an ideal elastic material, which is constantly deformed under the action of stress, and maintains its deformation until the stress is eliminated, so as to return to its original shape. This elongation or compression stress, the ratio of stress to strain is Young's modulus, namely:

$$E = (F/S)/(\Delta L/L)$$

Where E is Young's modulus, L is the primary length of cardiomyocytes, S is the surface area of cardiomyocytes, F is the ultrasonic radiation force, $\Delta L$ is the slight change in the length of cardiomyocytes after ultrasound.

After the deformation of cardiac myocytes, the membrane tension will also change[19][20][21] To facilitate the study of cell membrane tension, we idealize cardiac myocytes as standardized cylinders. In membrane stretching experiments, it was found that plasma membrane was separated from cytoskeleton during the formation of the tube[22][23], and only plasma membrane was found in the tube, whose shape was similar to that of cardiac muscle cells and could be idealized into a cylindrical shape. Based on such experimental results, We propose a quantitative model that can calculate the plasma membrane tension of cardiomyocytes under different shape variables (i.e., cylinder height variations).

$$\sigma = K/2R^2$$

$$L = (S_t - S_b)/2\pi R$$

$\sigma$ is the plasma membrane tension, K is the bending stiffness of membrane, R is the radius of cardiomyocyte, L is the length of cardiomyocyte, $S_t$ is the surface area of cardiomyocyte, $S_b$ is the base area of cardiomyocyte.

The number of phospholipid bilayer of cardiomyocytes remained constant during ultrasound, so we believed that the surface area of cardiomyocytes remained constant during stretching and compression.

It can be seen from the above two formulas that the plasma membrane tension of cardiomyocytes is related to the shape variables of cardiomyocytes, and the plasma membrane tension can be calculated under different shape variables. The initial values of the cardiomyocytes were assumed to be 150 $\mu m$ length and 35 $\mu m$ width. The standard value of cell membrane bending stiffness is 0.25PN $\mu m$ [24].The changes of plasma membrane tension of myocardial cell with stretching were shown in Table 1.

It is found that Piezo1 channels are more sensitive to plasma membrane tension than other stretch activated channels[25][26][27] .Piezo1 channel is stimulated by mechanical force to open the central hole of the channel, so that positively charged ions flow into the cell, thus triggering the electrochemical process in the cell. It is a key sensor of mechanical force in the cardiovascular system[28][29][30][31] . And Piezo1 is so abundant in the body that no immune response occurs when Piezo1 invades. In this paper, The SAC channel we consider is mainly Piezo1 channel.

The opening probability of Piezo1 channel is directly affected by the tension of the plasma membrane, and in the experiment [32] the Boltzmann relation between the Piezo1 opening probability $P_0$ and the plasma membrane tension $\sigma$ is obtained:

$$P_0 = 1/(1+\exp(-\frac{\sigma_1 - 5.1}{0.75}))$$

Thus, we obtained the relationship between ultrasonic radiation, myocardial cell length, plasma membrane tension and opening probability. See table 2.

## 3 RESULT

### 3.1 Regulation process of ultrasonic radiation force by hypercompression and hypertension

In order to verify that Piezo1 channel can be opened after myocardial cell deformation and plasma membrane tension change under ultrasonic radiation force, We selected a node in the myocardium tissue to simulate the piezo1 channel and observed the changes in plasma membrane tension, opening probability, length of myocardium cells with time and excitation state of myocardium cells,as shown in Figure 1 . It can be seen that myocardial cells have a certain opening probability during their own cycle contraction and diastole, but its value is relatively small, because of the periodic excitation, Piezo1 channel would be deactivated and the high probability of Piezo1 opening could not be reached by direct stretching. Therefore, the myocardium cells needed to be hypercompression first, so that Piezo1 channel was in an excited state and then stretched.

### 3.2 Sonogenetics removal of tachycardia, ventricular fibrillation

To verify our model, we simulated both global and local ultrasonic elimination of spiral waves and turbulence in a two-dimensional square myocardium. The isotropic medium was selected, the size of 10cm×10cm tissue was 500×500, and the boundary was zero flow boundary. The time step was 0.01ms, and the space step was 0.02cm.

A large number of piezo1 channels are opened under the action of ultrasonic radiation force in the whole region. The ions flowing through the Piezo1 channel generate a stimulus current large enough that the cardiomyocytes at each node are excited by the current to generate an action potential. Due to the effect of the refractory period during the action potential, the cardiomyocytes in the excited state cannot be excited again, and the spiral wave or turbulent propagation paths are cut off. Spiral wave or turbulence that cannot continue to propagate will slowly disappear with the change of action potential,as shown in Figure 2 and Figure 3.

The essence of the whole region ultrasonic radiation force is to make all cells in the myocardial tissue in an refractory state, so that the spiral wave or turbulence can not propagate. In principle, if the ultrasonic radiation force stimulates the response period in the local strip region, the spiral wave and turbulence cannot pass through or bypass the region to continue to propagate.If the strip region moves in one direction, the region that has just passed through remains in an refractory state for a period of time, during which time the transmembrane potential of the nodes where the spiral wave or turbulence cannot propagate gradually returns to a resting state. With the advantage of phased array ultrasound that can change the focusing position flexibly, we can sweep the ultrasonic radiation force bar region from one side of the tissue to the other side. In this way, the spiral wave or turbulence in the area scanned by ultrasound will disappear, and the transmembrane potential of myocardial cells will return to the normal resting state,as shown in Figure 4 and Figure 5.

**4       Discussion**

Piezo1 channel does not require a complete cytoskeleton [33][34]for gating, but it can also be activated by cytoskeleton forces and the cytoskeleton proteins themselves can change plasma membrane tension. This does not exclude the role of the cytoskeleton in Piezo1 as cortical tension was not considered in this paper due to lack of experimental results.Cardiomyocytes are anisotropic[35][36]. In this paper, isotropy is considered. If anisotropy is considered, the shape variables of each myocardial cell are different under the influence of anisotropy under ultrasonic radiation force. After the shape variables of cardiomyocytes are obtained according to the anisotropic characteristics, the plasma membrane tension and opening probability can be solved according to the method studied in this paper, so that the ultrasonic radiation force can regulate the opening probability of ion channels. The method studied in this paper can also be applied to nerve cells [37][38]by changing the idealized cylindrical structure of cardiac muscle cells into a tree-like structure.

**5       Conclusion**

In this paper, we studied and solved the problem of artificially regulating the opening probability of SAC channel in sonogenetics, and could accurately calculate the value of plasma membrane tension of cardiomyocytes after different deformation. In the experiment of treating arrhythmia with sonogenetics, the ultrasonic radiation force can activate the SAC channel in the focus area and change the excitability of myocardial cells in the area, so as to eliminate tachycardia (the corresponding electrocardiogram signal wave is spiral wave) and cardiac fibrillation (the corresponding electrocardiogram signal wave is turbulence). In this process, how ultrasound activates SAC channel in focusing region is a key problem. In order to solve this problem, we have conducted a series of research and exploration. After applying the ultrasonic radiation force to the myocardial tissue, the myocardial tissue will have certain deformation, and the cardiomyocytes will also have corresponding changes. When cardiomyocytes are stretched or compressed, their membrane tension changes accordingly. It is found that Piezo1 channel is mainly controlled by the plasma membrane tension. Therefore, we refer to the calculation method of plasma membrane tension in tube which is very similar to cardiomyocytes. During the formation of the tube, the plasma membrane is separated from the cytoskeleton, and only the plasma membrane is present in the tube. The shape of the tube is similar to that of cardiomyocytes, both of which can be idealized as cylinders. Based on this method, we can calculate the plasma membrane tension of cardiomyocytes under different shape variables. Then, according to the Boltzmann relationship between plasma membrane tension and the opening probability obtained in the experiment, the probability of Piezo1 channel opening caused by the change of plasma membrane tension after ultrasonic radiation force " hypercompression " and " hypertension " was obtained, thus changing the excitability of cardiomyocytes to treat arrhythmia.

## 6   Author Contributions

Jing-Jing Wang conducted simulation and drafted the manuscript. Xiang Gao put forward the concept, analyzed the data, and reviewed the manuscript.

## 7   Funding

This work is supported by the National Natural Science Foundation of China (Program No. 11727813),Natural Science Basic Research Program of Shaanxi (Program No.2019JQ-163 and NO.2019JQ-810), the Max Planck Society and the German Center for Cardiovascular Research.

## 8   Data Availability Statement

The original contributions presented in the study are included in the article, further inquiries can be directed to the corresponding author.

## 9   Reference styles

**Tables**

Table 1: Changes of plasma membrane tension of myocardial cell with stretching

|  | 1 | 2 | 3 | 4 | 5 | 6 |
|---|---|---|---|---|---|---|
| L/$\mu m$ | 155 | 167 | 180 | 195 | 212 | 232 |
| T/mN $m^{-1}$ | 4.1 | 4.4 | 4.7 | 5.0 | 5.4 | 5.8 |

Table 2: The relationship between ultrasonic radiation force, cardiomyocytes length, plasma membrane tension and opening probability

| Force/mN $m^{-1}$ | Length/$m * 10^{-6}$ | Plasma membrane tension/mN $m^{-1}$ | Opening probability |
|---|---|---|---|
| 2.2 | 67.0 | 2.4 | 0.03 |
| 0.0 | 126.0 | 3.5 | 0.11 |
| 0.0 | 155.0 | 4.1 | 0.21 |
| 12.5 | 290.0 | 7.2 | 0.95 |

**Figure captions**

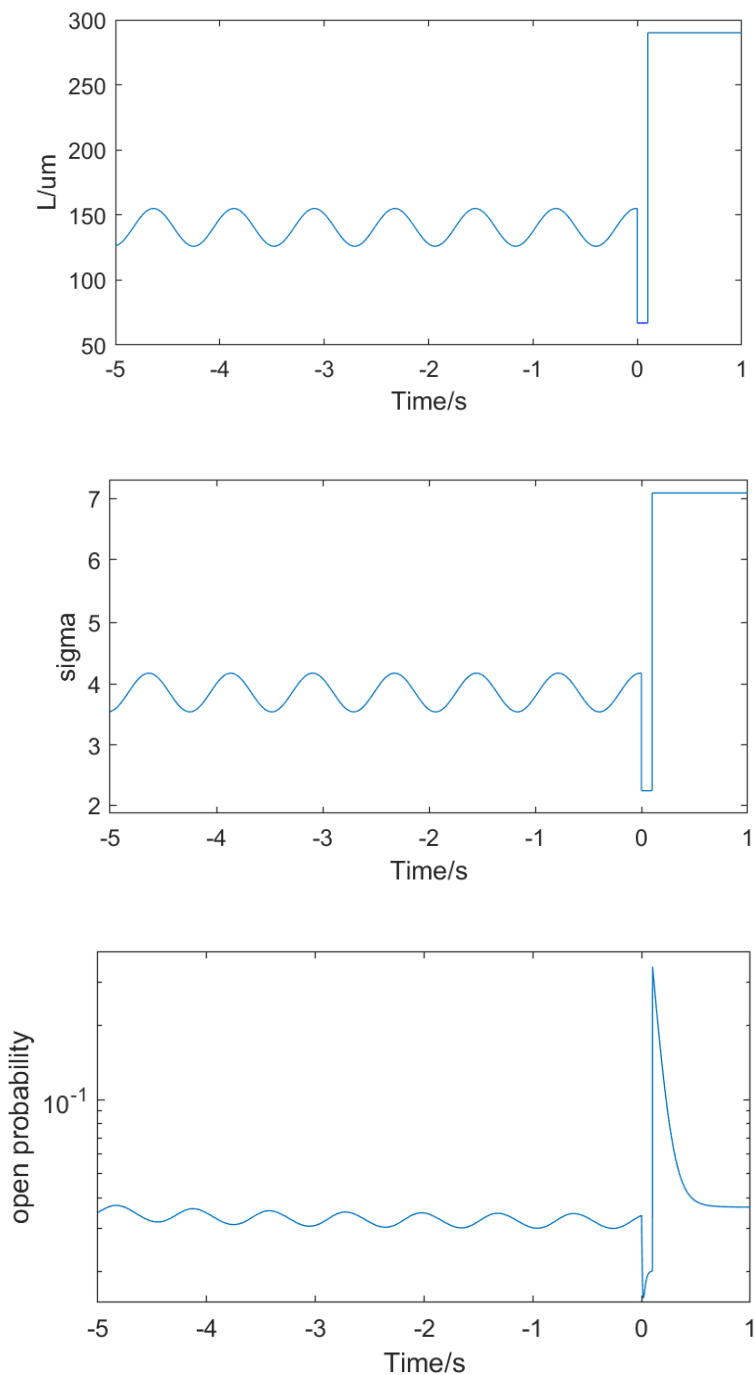

Figure1.Changes of length of cardiomyocytes, piezo1 and plasma membrane tension with time under ultrasonic radiation force. Before 0s, cardiomyocytes contracted and dilated cyclically. At 0s, the ultrasonic radiation force was added, which first hypercompression and then hypertension, and the probability of piezo1 and plasma membrane tension decreased first and then increased.

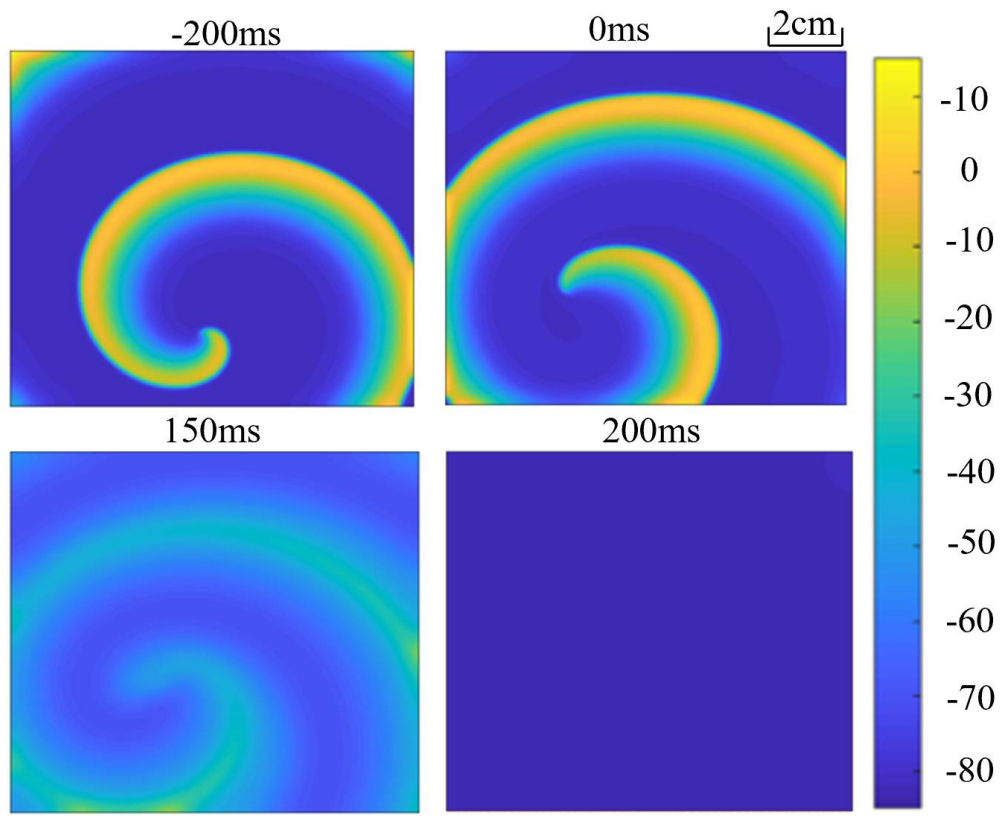

Figure2. The process of resetting spiral wave in whole region ultrasound. The yellow region has a high transmembrane potential, and the cardiomyocytes in this region are excited. The blue region has a lower transmembrane potential, indicating that the cardiomyocytes are in a resting state. A large number of piezo1 channels are opened under the action of ultrasonic radiation force in the whole region. The ions flowing through the Piezo1 channel generate a stimulus current large enough that the cardiomyocytes at each node are excited by the current to generate an action potential. Due to the effect of the refractory period during the action potential, the cardiomyocytes in the excited state cannot be excited again, and the spiral wave propagation path is cut off. The spiral wave that cannot continue to propagate will slowly disappear with the change of action potential.

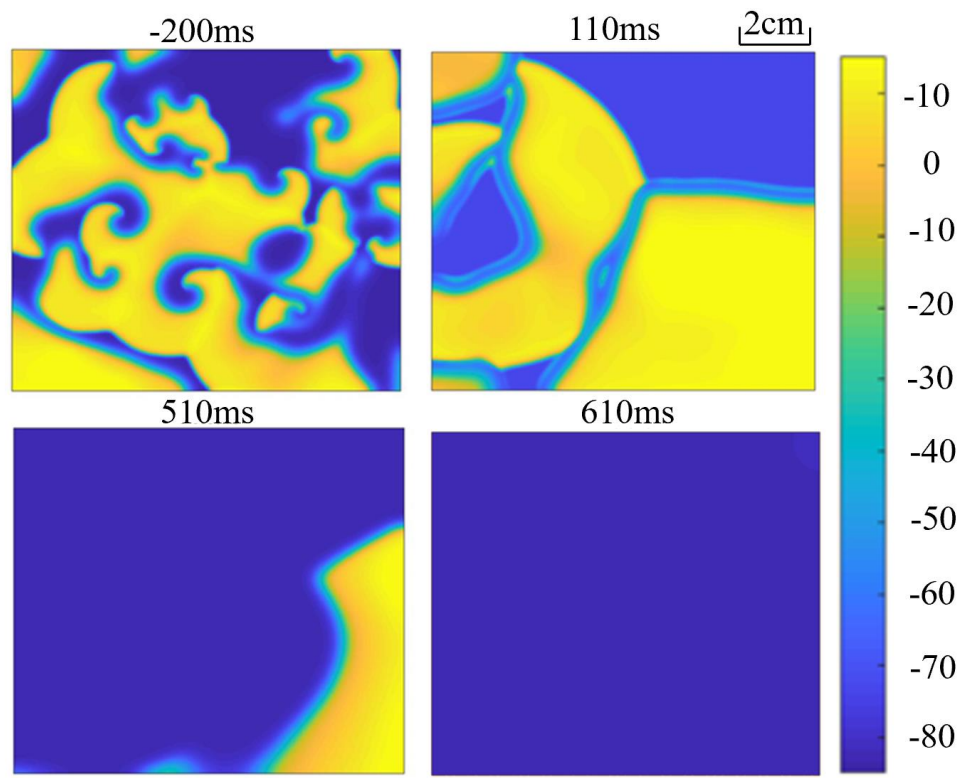

Figure3.The process of ultrasonic resetting turbulence in the whole region. The yellow region has a high transmembrane potential, and the cardiomyocytes in this region are excited. The blue region has a lower transmembrane potential, indicating that the cardiomyocytes are in a resting state. A large number of piezo1 channels are opened under the action of ultrasonic radiation force in the whole region. The ions flowing through the Piezo1 channel generate a stimulus current large enough that the cardiomyocytes at each node are excited by the current to generate an action potential. Due to the effect of the refractory period during the action potential, cardiomyocytes in the excited state cannot be excited again, and the path of turbulence propagation is cut off. Turbulence that cannot continue to propagate will slowly disappear with the change of action potential.

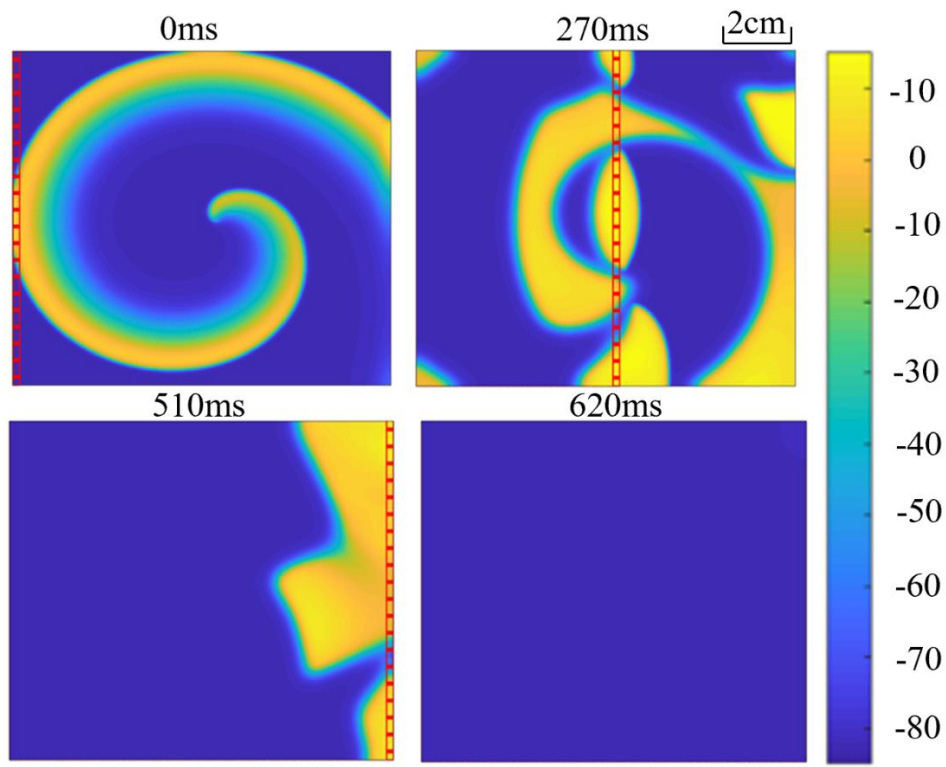

Figure4.The process of sweeping spiral wave in the area of strip refractory period produced by ultrasound. The yellow region has a high transmembrane potential, and the cardiomyocytes in this region are excited. The blue region has a lower transmembrane potential, indicating that the cardiomyocytes are in a resting state. The ultrasonic radiation force stimulates the refractory period in the local strip region, and the spiral wave cannot pass through or bypass the region to continue to propagate. If the strip region moves in one direction, the region that has just passed through remains in an irreversible state for some time, during which time the transmembrane potential of the node where the spiral wave cannot travel gradually returns to its resting state.

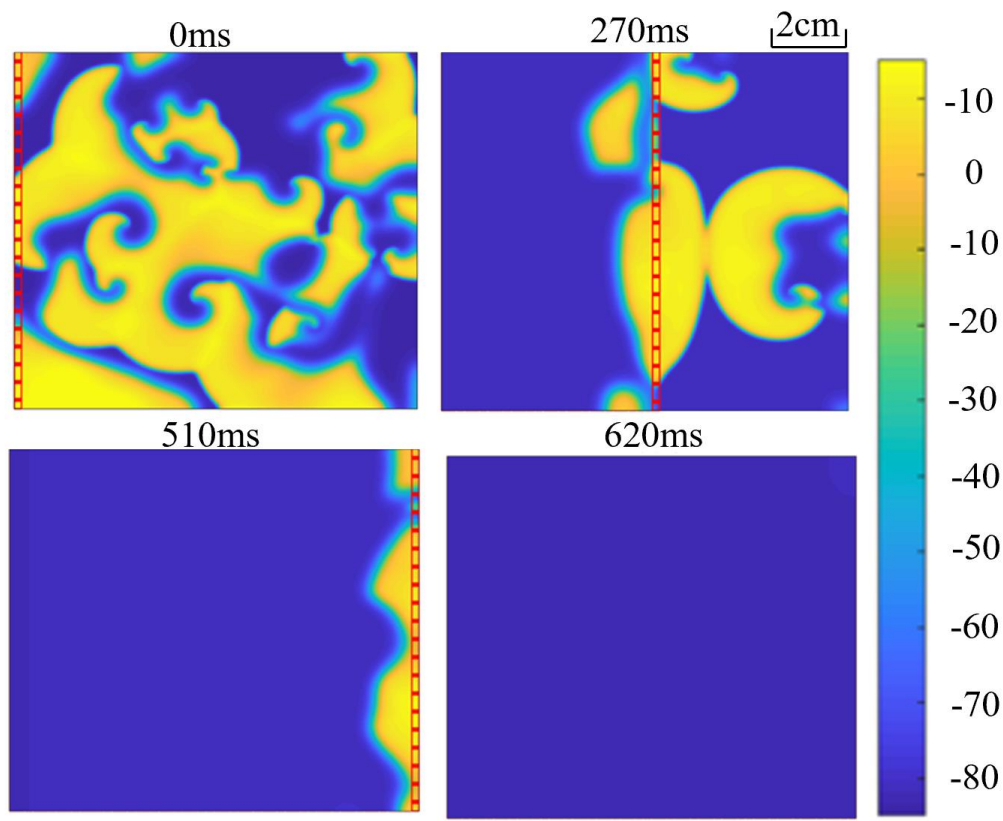

Figure5.The process of sweeping turbulence in the area of strip refractory period produced by ultrasound. The yellow region has a high transmembrane potential, and the cardiomyocytes in this region are excited. The blue region has a lower transmembrane potential, indicating that the cardiomyocytes are in a resting state. The ultrasonic radiation force stimulates the refractory period in the local strip region, and the turbulence cannot pass through or bypass the region to continue to propagate. If the strip region moves in one direction, the region that has just passed will remain in an refractory state for a period of time, during which time the transmembrane potential of the turbulence nodes that cannot propagate will gradually return to a resting state.

13